\documentclass[12pt]{article}
\usepackage{amsmath,amssymb}

\makeatletter
\@addtoreset{equation}{section}
\makeatother

\def\ben{\begin{equation}}
\def\een{\end{equation}}

  \let\n=\nu

\let\C=\Chi

\def\nn{\nonumber} \def\bd{\begin{document}} \def\ed{\end{document}}
\def\ds{\documentstyle} \let\fr=\frac \let\bl=\bigl \let\br=\bigr
\let\Br=\Bigr \let\Bl=\Bigl
\let\bm=\bibitem
\let\na=\nabla
\let\pa=\partial \let\ov=\overline
\newcommand{\be}{\begin{equation}}
\newcommand{\ee}{\end{equation}}
\def\ba{\begin{array}}
\def\ea{\end{array}}
\def\ft#1#2{{\textstyle{\frac{\scriptstyle #1}{\scriptstyle #2}}}}
\def\fft#1#2{\frac{#1}{#2}}

\def\del{\partial}
\def\vp{\varphi}
\def\sst#1{{\scriptscriptstyle #1}}
\def\oneone{\rlap 1\mkern4mu{\rm l}}
\def\td{\tilde}
\def\wtd{\widetilde}
\def\ie{\rm i.e.\ }
\def\dalemb#1#2{{\vbox{\hrule height .#2pt
        \hbox{\vrule width.#2pt height#1pt \kern#1pt
                \vrule width.#2pt}
        \hrule height.#2pt}}}
\def\square{\mathord{\dalemb{6.8}{7}\hbox{\hskip1pt}}}
\newcommand{\ho}[1]{$\, ^{#1}$}
\newcommand{\hoch}[1]{$\, ^{#1}$}
\newcommand{\bea}{\begin{eqnarray}}
\newcommand{\eea}{\end{eqnarray}}
\newcommand{\ra}{\rightarrow}
\newcommand{\lra}{\longrightarrow}
\newcommand{\Lra}{\Leftrightarrow}
\newcommand{\ap}{\alpha^\prime}
\newcommand{\bp}{\tilde \beta^\prime}
\newcommand{\tr}{{\rm tr} }
\newcommand{\Tr}{{\rm Tr} }
\def\0{{\sst{(0)}}}
\def\1{{\sst{(1)}}}
\def\2{{\sst{(2)}}}
\def\3{{\sst{(3)}}}
\def\4{{\sst{(4)}}}
\def\5{{\sst{(5)}}}
\def\6{{\sst{(6)}}}
\def\7{{\sst{(7)}}}
\def\8{{\sst{(8)}}}
\def\n{{\sst{(n)}}}
\def\cA{{{\cal A}}}
\def\cB{{{\cal B}}}
\def\cF{{{\cal F}}}
\def\tV{\widetilde V}
\def\tW{\widetilde W}
\def\tH{\widetilde H}
\def\tE{\widetilde E}
\def\tF{\widetilde F}
\def\tA{\widetilde A}
\def\im{{{\rm i}}}
\def\tY{{{\wtd Y}}}
\def\ep{{\epsilon}}
\def\vep{{\varepsilon}}
\def\R{\rlap{\rm I}\mkern3mu{\rm R}}
\def\bD{{{\bar D}}}

\def\R{\rlap{\rm I}\mkern3mu{\rm R}}
\def\bD{{{\bar D}}}
\def\R{{{\mathbb R}}}
\def\C{{{\mathbb C}}}
\def\H{{{\mathbb H}}}
\def\CP{{{\mathbb C}{\mathbb P}}}
\def\RP{{{\mathbb R}{\mathbb P}}}
\def\Z{{{\mathbb Z}}}
\def\bA{{{\mathbb A}}}
\def\bB{{{\mathbb B}}}
\def\bC{{{\mathbb C}}}
\def\bD{{{\mathbb D}}}
\def\bE{{{\mathbb E}}}
\def\bZ{{{\mathbb Z}}}
\def\Re{{{\frak{Re}}}}
\def\Im{{{\frak{Im}}}}
\def\cosec{{\,\hbox{cosec}\,}}
\def\Gm{{\Gamma_{\!\! -}}}
\def\Gp{{\Gamma_{\!\! +}}}
\def\stan{{standard }}
\def\nonstan{{supernumerary }}

\newcommand{\auth}{Justin F. V\'azquez-Poritz}

\begin{document}
\begin{flushright}

UCTP-112-04\ \ \ \ \
UK-04-14\\
August  2004\ \ \ \
\bf hep-th/0408144\\
\end{flushright}

\begin{center}
{\large {\bf The Nuts and Bolts of Brane Resolution}}

\vspace{10pt} \auth

\vspace{10pt} {\it Department of Physics and Astronomy,\\
University of Kentucky, Lexington, KY 40506}

\vspace{10pt} {\it Department of Physics,\\
University of Cincinnati, Cincinnati OH 45221-0011}

\vspace{15pt}

\underline{ABSTRACT}
\end{center}

We construct various non-singular $p$-branes on higher-dimensional
generalizations of Taub-NUT and Taub-BOLT instantons. Among other
solutions, these include $S^1$-wrapped D3-branes and M5-branes, as
well as deformed M2-branes. The resulting geometries smoothly
interpolate between product spaces which include Minkowski
elements of different dimensionality. The new solutions do not
preserve any supersymmetry.



\pagebreak \setcounter{page}{1}

\tableofcontents
\addtocontents{toc}{\protect\setcounter{tocdepth}{2}}
\newpage

\section{Introduction}

Regular $p$-brane solutions can be constructed by a process which
has been referred to as ``resolution via transgression''
\cite{cglptrans}. The first step is to replace the standard flat
transverse space by a smooth space of special holonomy, which is
Ricci-flat and has fewer covariantly constant spinors. Next, the
$p$-brane solution is deformed by additional flux such that the
Chern-Simons terms modify the equation of motion and/or Bianchi
identity of the field strength. If the new transverse space has a
non-collapsing $n$-cycle, then it can support a harmonic $n$-form
which is square integrable at short distance. The result is a
completely non-singular geometry \footnote{The latter portion of
this procedure was first applied to the heterotic 5-brane, for
which the singularity could be smoothed out by Yang-Mills fluxes
\cite{andy}. Curiously, the standard flat transverse space did not
need to be replaced.}.

This procedure was applied to the D3-brane, for which the
six-dimensional transverse space was replaced by the deformed
conifold \cite{klebstra,ganpol,gubser}. Since the new transverse
space has a non-collapsing 3-cycle, it supports a square
integrable three-form flux, which serves to resolve the
singularity. This singularity resolution procedure has been
applied to many other $p$-branes, including D2-branes on spaces of
$G_2$ holonomy \cite{cglptrans}, M2-branes on spaces of Spin(7)
holonomy \cite{cglpns2d2,cglptrans,cglpsten,beckers,deklm,
htr,becker,hk,cglphyper,cglpspin7,cgllp,gukov}, and others
\cite{herzog1,herzog2}, to give a small sample.

Since the standard flat transverse space is replaced by a space of
special holonomy, the supersymmetry is reduced to a minimum. In
fact, the main motivation for considering such solutions is
because they may constitute viable gravity duals of
strongly-coupled Yang-Mills field theories with reduced
supersymmetry. This indicates that they may shed light on
confinement and chiral-symmetry breaking. This was explored for
the case of the D3-brane and the dual ${\cal N}=1$, $D=4$
superconformal Yang-Mills theory in
\cite{kt,klebstra,ganpol,gubser}.

Previous work on brane resolution has incorporated transverse
spaces of special holonomy, in order to study dual gauge theories
of minimal supersymmetry. However, resolved brane solutions are
certainly of interest in their own right. Unlike typical brane
solutions which require a source term that is beyond supergravity,
resolved branes are complete purely within the framework of
supergravity. An interesting point is that the procedure of
resolving singularities does not depend on the presence of
supersymmetry. In this paper, we illustrate this point by
considering $p$-brane solutions on higher-dimensional
generalizations of Taub-NUT and Taub-BOLT instanton spaces
\cite{awad,lpp,cglpspin7} which do not preserve any supersymmetry.

\section{$S^1$ wrapped D3-brane}

The D3-brane of type IIB supergravity is supported by the
self-dual 5-form field strength, with a six-dimensional Ricci-flat
transverse space. Due to the Bianchi identity $dF_\5=F_\3^{\rm
NS}\wedge F_\3^{\rm RR}$, one can construct a fractional D3-brane
if the transverse space has a self-dual 3-cycle.  If instead the
transverse space has a 2-cycle $L_\2$, we can construct an
$S^1$-wrapped D3-brane with one of the world-volume coordinates
fibred over the transverse space \cite{lvres,wrapped}. The
solution is given by
\bea ds_{10}^2 &=& H^{-\ft12}\, \Big(-dt^2 + dx_1^2 + dx_2^2 +
(dx_3 +\cA_\1)^2\Big) + H^{\ft12}\, ds_6^2\,,\nn\\
F_\5&=&dt\wedge dx_1\wedge dx_2\wedge (dx_3+\cA_\1)\wedge dH^{-1}
-
{\ast_6 dH}\nn\\
&&+ m\, {\ast_6 L_\2}\wedge (dx_3+\cA_\1) +
dt\wedge dx_1\wedge dx_2\wedge L_\2\,,\nn\\
d\cA_\1&=&m\, L_\2\,,\label{wrapd3gen} \eea
where $L_\2$ is a harmonic 2-form in the transverse space of the
metric $ds_6^2$, and $\ast_6$ is the Hodge dual with respect to
$ds_6^2$.  The equations of motion are satisfied, provided that
\be \square H = -\ft12 m^2 L_\2^2\,,\label{lapmod} \ee
where $\square$ is the Laplacian in $ds_6^2$.

Note that one can also have a fibred time-like direction, in which
case the D3-brane solution is given by
\bea ds_{10}^2 &=&
H^{-1/2}\,\Big(-(dt+A_\1)^2+dx_1^2+dx_2^2+dx_3^2
\Big) +H^{1/2}\,ds_6^2\,,\nn\\
F_\5 &=& (dt+A_\1)\wedge dx_1\wedge dx_2\wedge dx_3\wedge
dH^{-1}+d^3x\wedge L_\2+{\rm dual\,\, terms}, \eea
which can be interpreted as a rotating D3-brane.

\subsection{On Taub-NUT/BOLT with ${\cal B}=\CP^2$}

The six-dimensional Taub-NUT/BOLT instanton metric with base space
$\CP^2$ is given by \cite{awad,cglpspin7}
\be ds_6^2 = F^{-1}\,dr^2+4N^2\,F\,(d\tau+A)^2+R^2\,d\Sigma_4^2\,,
\label{6Dnutcp} \ee
where $R^2=r^2-N^2$. $d\Sigma_4^2$ is the metric over $\CP^2$,
given by
\be d\Sigma_2^2=\fft{du^2}{V^2}+\fft{u^2}{4V^2} (d\psi+\cos\theta
\,d\phi)^2+\fft{u^2}{4V}(d\theta^2+\sin^2\theta\,d\phi^2)\,.
\label{cp2metric} \ee
\be A=\fft{u^2}{2V}(d\psi+\cos\theta\,d\phi)\,, \label{cp2A} \ee
and
\be V=1+u^2/6\,. \label{cp2V} \ee
The function $F$ is given by
\be F=\fft{\fft13 r^4-2N^2\,r^2-2M\,r-N^4}{R^4}\,. \label{F} \ee

For the appropriate periodicity of the fibre coordinate $\tau$, we
can avoid a conical singularity. In fact, for the base space
$\CP^2$, the six-dimensional Taub-NUT ($M=-\fft43 N^3$ and $r\ge
N$) and Taub-BOLT ($M=\fft43 N^3$ and $r\ge 3N$) are free of
singularities. At short distance, the Taub-NUT approaches $\R^8$
while the Taub-BOLT approaches $\R^2\times \CP^2$. At large
distance, they both are asymptotically cylindrical, having the
form of an $S^1$ bundle over a cone with the base $\CP^2$, which
we shall denote as $C(\CP^2)\ltimes S^1$.

We are interested in finding a harmonic 2-form supported by the
metric (\ref{6Dnutcp}). The most general ansatz for a 2-form with
respect to the isometry of (\ref{6Dnutcp}) is given by
\be L_\2 = u_1\, e^0\wedge e^5 + u_2\, e^1\wedge e^2 + u_3\,
e^3\wedge e^4 \,, \label{Lansatz} \ee
expressed in the vielbein basis
\bea e^0 &=& \fft{1}{\sqrt{F}}\,dr\,,\qquad
e^1=R\,\fft{u}{2\sqrt{V}}\, d\theta\,,\qquad
e^2=R\,\fft{u}{2\sqrt{V}}\,\sin\theta\,d\phi\,, \qquad
e^3=\fft{R}{V}\,du\,,\nn\\ e^4 &=& R\,\fft{u}{2V}\,
(d\psi+\cos\theta\,d\phi)\,,\qquad e^5=\sqrt{F}\,
[d\tau+\fft{u^2}{2V}\,(d\psi+\cos\theta\,d\phi)]\,. \eea
The closure and co-closure of $L_\2$ yield the following
solutions:
\be u_1=\Big( \fft{r+N}{r-N}\Big)^{\pm \fft{\sqrt{8+N^2}}{2N}}\,
(r^2-N^2)^{-3/2}\,,\qquad -u_2=u_3=\fft14 (r\mp \sqrt{8+N^2})\,
u_1\,. \label{u} \ee
It has a non-trivial flux and is not normalizable at large
distance. For the Taub-BOLT metric ($M=\fft43 N^3$ and $r\ge 3N$),
both two-forms are square integrable at short distance. On the
other hand, for the Taub-NUT metric ($M=-\fft43 N^3$ and $r\ge N$)
only the two-form with the negative sign is square integrable at
short distance. We solve for the corresponding $H$ in the case
$N=1$.

First, we consider the Taub-NUT metric. For the top sign in
(\ref{u}), there are only singular solutions. This can be seen a
priori, since the corresponding two-form is not square integrable.
On the other hand, for the bottom sign in (\ref{u}), there is a
regular solution to $H$. For this first example, we will show
explicit details. In this case, (\ref{lapmod}) can be written as
\be \fft{1}{(r-1)^2}\,\partial_r \,(r-1)^3\,(r+3)\,
\partial_r H= -\fft{3m^2\,(r^2+6r+17)}{8(r+1)^4}\,. \ee
The solution is given by
\be H=1+\fft{3m^2}{16(r+1)^2}+\Big( c-\fft{9}{512}m^2\Big) \Big[
\fft{4}{r-1}-\fft{8}{(r-1)^2}+{\rm log} \Big( \fft{r-1}{r+3}\Big)
\Big]\,. \ee
Choosing the integration constant $c=\fft{9}{512}m^2$ yields a
regular solution given by
\be H=1+\fft{3m^2}{16(r+1)^2}\,. \ee
For future examples given in this paper, we will only present the
regular solution that results from the appropriate choice of the
integration constant.

At both short and large distances, $H$ asymptotes to a constant.
Therefore, the D3-brane geometry smoothly interpolates between
$M_3\times \R^6 \ltimes S^1$ (a product space of three-dimensional
Minkowski spacetime and a $U(1)$ bundle over $\R^6$) at short
distance to $M_4\times C(\CP^2)\ltimes S^1$ (a product space of
four-dimensional Minkowski spacetime and a $U(1)$ bundle over a
cone with base $\CP^2$) at large distance.

Next, we turn to the Taub-BOLT metric. In this case, since the
two-form is square integrable for both signs in (\ref{u}), both of
the corresponding solutions are regular. These are given by
\be H=1+\fft{3m^2\,(\pm 6+5r\pm 25r^2+10r^3)}{160(r\pm 1)^5}\,.
\ee
As before, $H$ is asymptotically constant at short and large
distances. The D3-brane geometry smoothly goes from $M_3\times
(\R^2\times \CP^2)\ltimes S^1$ at short distance to $M_4\times
C(\CP^2)\ltimes S^1$ at large distance.

Since the Taub-BOLT instanton supports two independent harmonic
two-forms, these can be superimposed to form a more general
$S^1$-wrapped D3-brane. It is also possible to have a regular
$T^2$-wrapped D3-brane, for which two of the worldvolume
directions are fibred over the two two-forms respectively. The
resulting short distance geometry of the D3-brane would then be
$M_2\times (\R^2\times \CP^2)\ltimes T^2$.

\subsection{On Taub-BOLT with ${\cal B}=S^2\times S^2$}

The six-dimensional Taub-NUT/BOLT instanton metric with base space
$S^2\times S^2$ is given by \cite{awad}
\be ds_6^2 = F^{-1}\,dr^2+4N^2\,F\,\sigma^2+(r^2-N^2) (d\Omega_2^2
+d{\td \Omega}_2^2)\,, \label{6Dnut} \ee
where
\bea d\Omega_2^2 &=& d\theta^2+\sin^2\theta\,d\phi^2\,,\qquad
d{\td \Omega_2^2}=d{\td \theta}^2+\sin^2{\td \theta}\,d{\td
\phi}^2\,,\nn\\
\sigma &=& d\psi+\cos\theta\,d\phi+\cos {\td \theta}\,d{\td
\phi}\,, \eea
The function $F$ is given by (\ref{F}). For the base space
$S^2\times S^2$, the Taub-NUT has a curvature singularity.
However, the Taub-BOLT ($M=\fft43 N^3$ and $r\ge 3N$) is
completely smooth, and goes from $\R^2\times S^2\times S^2$ at
short distance to $C(S^2\times S^2)\ltimes S^1$ at large distance.

The most general ansatz for a 2-form with respect to the isometry
of (\ref{6Dnut}) is given by (\ref{Lansatz}) expressed in the
vielbein basis $e^0=1/\sqrt{F}\, dr$, $e^1=R\,d\theta$,
$e^2=R\,\sin\theta\,d\phi$, $e^3=R\, d{\td \theta}$, $e^4=R\,\sin
{\td \theta}\,d{\td \phi}$ and $e^5=2N \sqrt{F}\, \sigma$. The
closure and co-closure of $L_\2$ yield the following solutions:
\be u_1=-\fft{4N}{(r\pm N)^3}\,,\qquad u_2=u_3 =\fft{r\pm
3N}{(r\pm N )^3}\,. \ee
It has a non-trivial flux.  The square of this form is
\be L_\2^2=\fft{32N^2+4(r\pm 3N)^2}{(r\pm N)^6} \,. \ee
Thus, it is square integrable for $r\rightarrow 0$ but not
normalizable at large distance.

For the negative sign, for an appropriate choice of integration
constant, there is a regular solution to $H$ in (\ref{lapmod})
given by
\be H=1+\fft{m^2}{(r-N)^2}\,. \ee
For the positive sign, an appropriate choice for the integration
constant leads to the regular solution
\be H=1+\fft{(6N^3+5N^2\,r+25N\,r^2+10r^3)\,m^2}{10(N+r)^5}\,. \ee
All of these geometries smoothly interpolate from $M_3\times
(\R^2\times S^2\times S^2)\ltimes S^1$ at short distance to
$M_4\times C(S^2\times S^2)\ltimes S^1$ at large distance. Again,
we can superimpose the above solutions to get a regular
$T^2$-wrapped D3-brane.

However, Taub-BOLT instantons with ${\cal B}=S^2\times S^2..\times
S^2$ do not admit a spin structure \cite{chamblinads,awad}. This
is because each $S^2$ factor generates an element of $H_2$ of odd
self-intersection. This applies to the $S^1$-wrapped D3-brane on a
Taub-BOLT with ${\cal B}=S^2\times S^2$ of this section, as well
as the deformed M2-brane on a Taub-BOLT with ${\cal B}=S^2\times
S^2\times S^2$ and the deformed or $S^1$-wrapped 5-brane on a
Taub-BOLT with ${\cal B}=S^2$, which we will discuss shortly.
Nevertheless, these spacetimes may still admit a $Spin^C$
structure\footnote{The author thanks Andrew Chamblin for
clarifying this point.}.

\subsection{On generalized Taub-NUT/BOLT}

\subsubsection{Taub-BOLT}

Recently a family of Taub-NUT and Taub-BOLT metrics were found
which have an additional spherical element \cite{lpp}. Included is
a six-dimensional Taub-BOLT solution with the topology $S^2$ times
a $\R^2$ bundle over the base space $S^2$. The metric is given by
\cite{lpp}
\be ds_6^2 = F^{-1}\,dr^2+4F\,\sigma^2+(r^2-1)
d\Omega_2^2+r^2\,d{\td \Omega}_2^2\,, \label{alt6Dbolt} \ee
where
\bea d\Omega_2^2 &=& d\theta^2+\sin^2\theta\,d\phi^2\,,\qquad
d{\td \Omega_2^2}=d{\td \theta}^2+\sin^2{\td \theta}\,d{\td
\phi}^2\,,\nn\\
\sigma &=& d\psi+\cos\theta\,d\phi\,, \eea
The function $F$ is given by
\be F=\fft{(r+1)(r-2)}{3r(r-1)}\,. \ee
Also, $r>2$. This Taub-BOLT geometry goes from $\R^2\times
S^2\times S^2$ at short distance to $C(S^2\times S^2)\ltimes S^1$
at large distance.

     We are interested in finding a harmonic 2-form supported by this
metric. The most general ansatz for a 2-form with respect to the
isometry of (\ref{alt6Dbolt}) is given by (\ref{Lansatz})
expressed in the vielbein basis $e^0=1/\sqrt{F}\, dr$,
$e^1=\sqrt{r^2-1}\,d\theta$,
$e^2=\sqrt{r^2-1}\,\sin\theta\,d\phi$, $e^3=r\, d{\td \theta}$,
$e^4=r\,\sin {\td \theta}\,d{\td \phi}$ and $e^5=2\sqrt{F}\,
\sigma$. The closure and co-closure of $L_\2$ yield three
solutions. The first one is given by
\be u_1=\fft{3r^2-1}{r^2\,(r^2-1)^2}\,,\qquad
u_2=\fft{2}{r(r^2-1)^2}\,,\qquad u_3=0\,, \ee
corresponding to the regular $H$ given by
\be H=\fft{m^2\,(9+20r+9r^2-12r^3-8r^4)}{6r(r-1)(r+1)^3} +\fft43
m^2\,{\rm log} \Big( \fft{r+1}{r}\Big)\,. \ee
The second solution is given by
\be u_1=\fft{r}{(r^2-1)^2}\,,\qquad u_2=\fft{3-r^2}{2(r^2-1)^2}\,,
\qquad u_3=0\,, \ee
with
\be H=1+\fft{m^2\,(11+18r-3r^2-8r^3)}{24(r-1)(r+1)^3} +\fft13
m^2\, {\rm log} \Big( \fft{r+1}{r}\Big)\,. \ee
The third solution is
\be u_1=u_2=0\,,\qquad u_3=\fft{1}{r^2}\,, \label{thirdL} \ee
with
\be H=1-\fft{3m^2\,(1+4r)}{2r(1+r)}+6m^2\,{\rm log} \Big(
\fft{r+1}{r}\Big)\,. \ee
Notice that the second and third harmonic two-forms have
non-trivial flux and are not normalizable at large distance.
However, all three two-forms are square integrable for
$r\rightarrow 2$. All of the corresponding D3-brane geometries
asymptotically approach $M_4\times C(S^2\times S^2)\ltimes S^1$ at
large distance. However, they have different bundle structures at
short distance. The first two solutions are $M_3\times S^2\times
(\R^2\times S^2)\ltimes S^1$ at short distance, while the third is
$M_5\times S^2\times S^2 \ltimes S^1$. These solutions can be
superimposed to get, for example, a $T^3$-wrapped D3-brane.

\subsubsection{Taub-NUT}

Simply taking $r\rightarrow -r$ transforms the above
six-dimensional Taub-BOLT metric into a Taub-NUT metric, where now
$r\ge 1$ \cite{lpp}. This Taub-NUT metric is given by
(\ref{alt6Dbolt}) with
\be F=\fft{(r-1)(r+2)}{3r(r+1)}\,. \ee
Thus, the same six-dimensional local metric form extends smoothly
onto two different manifolds. This Taub-NUT geometry runs from
$\R^4\times S^2$ to $C(S^2\times S^2)\ltimes S^1$.

The three harmonic two-forms supported by these Taub-BOLT and
Taub-NUT metrics are identical, since they do not depend on the
function $F$. However, the resulting $H$ for each case are not
related by taking $r\rightarrow -r$, and must be solved from
scratch. In fact, only for the third harmonic two-form
(\ref{thirdL}) does there exist a regular solution, given by
\be H=1+\fft{3m^2}{2r}-\fft34 m^2\,{\rm log} \Big( \fft{r+2}{r}
\Big)\,. \ee
The D3-brane geometry smoothly runs from $M_3\times (\R^4\times
S^2)\ltimes S^1$ to $M_4\times C(S^2\times S^2)\ltimes S^1$.

\subsection{On Schwarzchild instanton}

We will now consider the $S^1$-wrapped D3-brane solution given by
(\ref{wrapd3gen}) for which the six-dimensional transverse space
is a Schwarzchild instanton, whose metric is given by
\be ds_6^2=f\,dx^2+f^{-1}\,dr^2+r^2\,d\Omega_4^2\,, \ee
where
\be f=1-\fft{M}{r^3}\,, \ee
and $r^3\ge M$. This geometry runs from $\R^2\times S^4$ at short
distance to $\R^6$ at large distance.

This metric supports a harmonic two-form
\be L_\2=\fft{1}{r^4}\,dx\wedge dr\,, \ee
which is normalizable at large distance and square integrable as
$r\rightarrow M^{1/3}$. The corresponding regular solution to $H$
in (\ref{lapmod}) is given by
\be H=1+\fft{m^2}{9M\,r^3}\,. \ee
This D3-brane geometry smoothly interpolates between $M_3\times
S^4\times \R^2\ltimes S^1$ and $M_{10}$.

\section{Deformed M2-brane}

The M2-brane of eleven-dimensional supergravity is supported by
the 4-form field strength, with an eight-dimensional Ricci-flat
transverse space. Due to the equation of motion $d\ast F_\4=\ft12
F_\4\wedge F_\4$, one can construct a resolved M2-brane if the
transverse space has a (anti)-self-dual 4-cycle. This type of
modification to the M2-brane, which makes use of the interaction
in $d\ast F_\4=\ft12 F_\4 \wedge F_\4$, has been greatly studied,
for example in \cite{cglpns2d2,cglptrans,cglpsten,beckers,deklm,
htr,becker,hk,cglphyper,cglpspin7,cgllp,gukov,m2}. The deformed
M2-brane is given by
\bea ds_{11}^2 &=& H^{-2/3}dx^{\mu}dx^{\nu}\eta_{\mu\nu}+H^{1/3}
ds_8^2\,,\nn\\
F_\4 &=& d^3x\wedge dH^{-1}+m\,G_\4\,, \label{deformed} \eea
where $G_\4$ is a harmonic self-dual 4-form in the Ricci-flat
transverse space $ds_8^2$. The equations of motion are satisfied,
provided that
\be \square H=-\fft{1}{48} m^2\,G_\4^2\,, \label{Heqn} \ee
where $\square$ is the Laplacian on $ds_8^2$.

Before discussing specific examples, we would like to mention that
the deformed M2-brane can be dimensionally reduced along the
worldvolume to give a deformed NS-NS string in type IIA theory,
given by
\bea ds_{10}^2 &=& H^{-3/4}\,(-dt^2+dx^2)
 +H^{1/4} ds_8^2\,,\nn\\
F_\4 &=& m\,G_\4\,,\qquad F_\3=dt\wedge dx\wedge dH^{-1}\,,\qquad
e^{2\phi}=H\,. \eea
This solution can be T-dualized to a regular type IIB pp-wave
given by
\bea ds_{10}^2 &=& -H^{-1}\,dt^2+H\,\Big(
dx+(H^{-1}-1)\,dt \Big)^2+ds_8^2\,,\nn\\
F_\5 &=& m_4\,\Big( dx+(H^{-1}-1)\,dt\Big) \wedge (G_\4+\ast_8
G_\4)\,. \eea

\subsection{On Taub-NUT/BOLT with ${\cal B}=\CP^3$}

For the transverse space, we will consider an eight-dimensional
Taub-NUT/BOLT instanton, whose metric is given by \cite{cglpspin7}
\be ds_8^2 = F^{-1}\,dr^2 + N^2\,F\,(d\tau+A)^2 + R^2\,
d\Sigma_6^2\,, \label{cp3metric} \ee
where
\be F=\fft{8(r^6-5N^2\,r^4+15N^4\,r^2-10M\,r+5N^6)}{5(r^2-N^2)^3}
\label{F8} \,, \ee
and $R^2=r^2-N^2$. $d\Sigma_6^2$ is the metric for $\CP^3$ given
by
\be d\Sigma_6^2=d\xi^2+\fft14 c^2\,d\Omega_2^2+\fft14
s^2\xi\,d{\td\Omega}_2^2+\fft14 s^2\xi\,c^2\xi\, \sigma^2\,, \ee
where
\bea d\Omega_2^2 &=& d\theta^2+\sin^2\theta\,d\phi^2\,,\qquad
d{\td\Omega}_2^2 = d{\td\theta}^2+\sin^2 {\td\theta}\,
d{\td\phi}^2\,,\nn\\
\sigma &=& d\psi-\cos\theta\,d\phi+\cos {\td\theta}\,d{\td\phi}\,,
\eea
and $c=\cos\xi$ and $s=\sin\xi$. Also,
\be A=\fft14 (c^2-s^2)\,d\psi-\fft12 c^2\,\cos\theta\,d\phi
-\fft12 s^2 \cos {\td\theta}\,d{\td\phi}\,. \ee
For the appropriate periodicity in the fibre coordinate $\tau$,
there is no conical singularity. At short distance, the Taub-NUT
geometry is $\R^8$ while the Taub-BOLT geometry is $\R^2\times
\CP^3$; both geometries asymptotically approach $C(\CP^3)\ltimes
S^1$ at large distance.

The veilbein for the 8-space described by (\ref{cp3metric}) are
given by
\bea e^0 &=& \fft{1}{\sqrt{F}}\, dr\,, \qquad
e^1=\fft{R}{2}c\,d\theta\,,\qquad e^2=
\fft{R}{2}c\,\sin\theta\,d\phi\,,\nn\\
e^3 &=& \fft{R}{2}s\,d{\td\theta}\,,\qquad e^4=\fft{R}{2}s\,\sin
{\td\theta}\,d{\td\phi}\,,\qquad e^5=R\,d\xi\,,\nn\\
e^6 &=& \fft{R}{2}sc\,(d\psi-\cos\theta\,d\phi+\cos
{\td\theta}\,d{\td\phi})\,,\qquad e^7 = N\,\sqrt{F}\,(d\tau+A)\,.
\eea

An (anti)self-dual 4-form on this 8-space is given by
\bea G_\4 &=& u_1^{\pm}\,(e^0\wedge e^7\wedge e^1\wedge e^2\pm
e^3\wedge e^4\wedge e^5\wedge e^6)\nn\\ && +u_2^{\pm}\,(e^0\wedge
e^7\wedge e^3\wedge e^4 \pm e^1\wedge e^2\wedge e^5\wedge
e^6)\nn\\ && +u_3^{\pm}\,(e^0\wedge e^7\wedge e^5\wedge e^6\pm
e^1\wedge e^2\wedge e^3\wedge e^4)\,. \label{G4} \eea
The closure of $G_\4$ yields the (anti)self-dual solution
\be u_1^{\pm}=u_2^{\pm}=-u_3^{\pm}=\fft{1}{(r\mp N)^4}\,, \ee
Both of these two-forms are normalizable at large distance. In the
case of the Taub-NUT ($M=\fft85 N^5$ and $r\ge N$), only the anti
self-dual two-form is integrable at short distance. This has the
corresponding regular solution for $H$ given by
\be H=1+\fft{5m^2\,(2N^2+3N\,r+3r^2)}{256N^5\,(r+N)^3}
+\fft{15m^2}{512N^6}\,{\rm arctan} \Big(
\fft{(r+N)(r+3N)}{2N\,(r+2N)}\Big)\,. \ee
The M2-brane geometry smoothly interpolates from $M_11$ at short
distance to $M_3\times C(\CP^3)\ltimes S^1$ at large distance.

Both the anti self-dual and the self-dual two-forms are square
integrable at short distance for the Taub-BOLT, which indicates
that the there exists corresponding regular solutions for $H$.
However, we have been unable to express these in a closed
analytical form. We expect that such M2-brane geometries would
smoothly run from $M_5\times \CP^3$ at short distance to
$M_3\times C(\CP^3)\ltimes S^1$ at large distance.

\subsection{On Taub-BOLT with ${\cal B}=S^2\times S^2\times S^2$
or $\CP^2\times S^2$}

We can replace the base space $\CP^3$ of the Taub-NUT/BOLT metric
(\ref{cp3metric}) by any other six-dimensional Einstein-K\"{a}hler
space, such as $S^2\times S^2\times S^2$ or $\CP^2\times S^2$
\cite{awad}. Since, in these cases, only the Taub-BOLT is free of
singularities, we will not consider the Taub-NUT in this section.

In the case of ${\cal B}=S^2\times S^2\times S^2$, the
corresponding $G_\4^2$ is the same as in the previous section.
This indicates that there is a corresponding regular solution for
$H$ though, again, there does not seem to be a closed form
analytical expression for it. Presumably this M2-brane geometry
would run from $M_5\times S^2\times S^2\times S^2$ to $M_3\times
C(S^2\times S^2\times S^2)\ltimes S^1$. However, as previously
mentioned, Taub-BOLT instantons of this topology do not admit a
spin structure, though they may still admit a $Spin^C$ structure
\cite{awad}.

We will now consider ${\cal B}=\CP^2\times S^2$ explicitly, though
we restrict ourselves to $N=1$ for simplicity. In this case, the
Taub-NUT/BOLT metric can be written as
\be ds_8^2 = F^{-1}\,dr^2 + 4F\,(d\tau+A)^2 + R^2\,
(d\Sigma_4^2+d\Omega_2^2)\,, \label{cp8metric} \ee
where $F$ is given in (\ref{F8}) divided by a factor of 8, and
$R^2=r^2-1$. $d\Sigma_4^2$ is the metric for $\CP^2$ given by
(\ref{cp2metric}) and (\ref{cp2V}). Also, $A$ is given by
(\ref{cp2A}), and $d\Omega_2^2=d\theta^2+\sin^2\theta\,d\phi^2$.
This Taub-BOLT geometry goes from $\R^2\times S^2\times \CP^2$ to
$C(S^2\times \CP^2)\ltimes S^1$.

An (anti)self-dual 4-form on this 8-space can be written as
(\ref{G4}), in the veilbein basis
\bea e^0 &=& \fft{1}{\sqrt{F}}\, dr\,, \qquad e^1=
\fft{u\,R}{2\sqrt{V}}\,d\theta\,,\qquad e^2=
\fft{u\,R}{2\sqrt{V}}\,\sin\theta\,d\phi\,,\nn\\
e^3 &=& \fft{R}{V}\, du\,,\qquad
e^4=\fft{u\,R}{2V}\,(d\psi+\cos\theta\,d\phi)\,,\qquad
e^5=R\,d{\td \theta}\,,\nn\\
e^6 &=& R\,\sin {\td\theta}\,d{\td\phi}\,,\qquad e^7 =
2\,\sqrt{F}\,(d\tau+A)\,. \eea
We will restrict ourselves to the case $N=1$.

The closure of $G_\4$ yields the following self-dual solutions:
\be u_1=-u_2=\fft{1}{2(r+1)^4}\,,\qquad u_3=\fft{1}{(r+1)^4}\,,
\label{1form} \ee
and
\be u_1=-u_2=\fft{1-9r+3r^2-3r^3}{24(r+1)^4\,(r-1)^3}\,,\qquad
u_3=\fft{3r^2+1}{3(r+1)^4\,(r-1)^3}\,. \ee
There are also two anti self-dual solutions given by
\be u_1=-u_2=-\fft{1}{4(r+1)^3\,(r-1)}\,,\qquad
u_3=\fft{1}{(r+1)^3\,(r-1)}\,, \ee
and
\be u_1=-u_2=\fft{1+9r+3r^2+3r^3}{24(r+1)^3\,(r-1)^4}\,,\qquad
u_3=\fft{3r^2+1}{3(r+1)^3\,(r-1)^4}\,. \ee
All of these two-forms are normalizable at large distance. For the
Taub-BOLT, all of them are also square integrable at short
distance. This indicates that each two-form has a corresponding
regular solution for $H$, even though we cannot express it in
closed analytical form. These M2-brane geometries would smoothly
go from $M_5\times S^2\times \CP^2$ to $M_3\times C(S^2\times
\CP^2)\ltimes S^1$.

\subsection{On generalized Taub-NUT}

For the transverse space, we will now consider a generalized
Taub-NUT instanton, whose metric is given by \cite{lpp}
\be ds_8^2 = \fft{5}{U}\,dr^2 + \fft{4U}{5}\,(\tau+A)^2 +
(r^2-1)\,d\Sigma_4^2+r^2\,d\Omega_2^2\,, \label{gen8metric} \ee
where
\be U=1-\fft{4(r^3-3r+2)}{3r\,(r^2-1)^2}\,. \ee
$d\Sigma_4^2$ is the metric for $\CP^2$ given by (\ref{cp2metric})
and (\ref{cp2V}). Also, $A$ is given by (\ref{cp2A}). This
Taub-NUT geometry goes from $\R^6\times S^2$ to $C(S^2\times
\CP^2)\ltimes S^1$.

The veilbein for the 8-space described by (\ref{gen8metric}) are
given by
\bea e^0 &=& \sqrt{\fft{5}{U}}\, dr\,, \qquad e^1=
\fft{u\,R}{2\sqrt{V}}\,d\theta\,,\qquad e^2=
\fft{u\,R}{2\sqrt{V}}\,\sin\theta\,d\phi\,,\nn\\
e^3 &=& \fft{R}{V}\, du\,,\qquad
e^4=\fft{u\,R}{2V}\,(d\psi+\cos\theta\,d\phi)\,,\qquad
e^5=r\,d{\td \theta}\,,\nn\\
e^6 &=& r\,\sin {\td\theta}\,d{\td\phi}\,,\qquad e^7 =
2\sqrt{\fft{U}{5}}\,(d\tau+A)\,. \eea

An (anti)self-dual 4-form on this 8-space is given by
\bea G_\4 &=& u_1\,(e^0\wedge e^7\wedge e^1\wedge e^2\pm e^3\wedge
e^4\wedge e^5\wedge e^6)\nn\\ && +u_2\,(e^0\wedge e^7\wedge
e^3\wedge e^4 \pm e^1\wedge e^2\wedge e^5\wedge e^6)\nn\\ &&
+u_3\,(e^0\wedge e^7\wedge e^5\wedge e^6\pm e^1\wedge e^2\wedge
e^3\wedge e^4)\,. \eea
The closure of $G_\4$ yields the following solutions:
\be u_1=\pm\fft{3r+1}{2r^2\,(r+1)^3}=-u_2\,,\qquad
u_3=\fft{1}{r\,(r+1)^3} \label{cp24form} \,, \ee
and
\be u_1=\pm\fft{1-10r^2-15r^4}{2r^2\,(r^2-1)^5}=-u_2\,,\qquad
u_3=\fft{3r^2+1}{r\,(r^2-1)^3}\,. \ee
All of these four-forms are normalizable at large distance.
However, only the (anti)self-dual pair given in (\ref{cp24form})
are square integrable as $r\rightarrow 1$ and, thus, have a
corresponding regular $H$ given by
\bea H &=& 1+\fft{15m^2}{16r}+\fft{45m^2}{8(r+1)}
-\fft{15m^2}{8(r+1)^2}+\fft{207}{128}\sqrt{15}\,m^2\, {\rm arctan}
\Big[ \sqrt{\fft35} (3+2r)\Big]\nn\\ && +\fft{15}{256}m^2\, {\rm
log} \Big[ \fft{r^{34}}{(r+1)^{16}\,(r^2+3r+8/3)^{9}}\Big] \,.
\eea
The geometry of this M2-brane interpolates from $M_9\times S^2$ at
short distance to $M_3\times C(S^2\times \CP^2)\ltimes S^1$ at
large distance.

\section{Deformed and $S^1$-wrapped 5-brane}

\subsection{Deformed heterotic 5-brane}

The deformed heterotic 5-brane is given by \cite{cglptrans}
\bea ds_{10}^2 &=& H^{-1/4}\,dx_{\mu}^2+H^{3/4}\, ds_4^2\,,\nn\\
e^{-\phi}\ast F_\3 &=& d^6x\wedge dH^{-1}\,,\qquad \phi=\fft12
{\rm log} H\,,\qquad F_\2=m\,L_\2\,, \eea
where $\ast$ is the Hodge dual with respect to $ds_{10}^2$ and
$L_\2$ is a harmonic two-form on $ds_4^2$. The equations of motion
are satisfied, provided that $L_\2$ is a self-dual two-form and
\be \square H=-\fft14 m^2\,L_\2^2\,, \label{eqnH5} \ee
where $\square$ is the Laplacian on $ds_4^2$. Note that an
overlapping 5-brane configuration can also be resolved
\cite{wrapped}.

We will consider the case in which $ds_4^2$ is the metric for the
Taub-NUT instanton, given by
\be ds_4^2 = F^{-1}\,dr^2+F\,(d\psi-2N\,\cos\theta\,d\phi)^2
+(r^2-N^2)(d\theta^2+\sin^2\theta\,d\phi^2)\,, \label{4Dnut} \ee
where
\be F=\fft{r^2-2M\,r+N^2}{r^2-N^2}\,. \ee

Harmonic 2-forms supported by this metric are given by
\be L_\2^{\pm} = \fft{2}{(r\pm N)^2}\, (e^0\wedge e^3 \pm
e^1\wedge e^2)\,, \ee
expressed in the vielbein basis $e^0=\fft{1}{\sqrt{F}}\, dr$,
$e^1=\sqrt{r^2-N^2}\,d\theta$,
$e^2=\sqrt{r^2-N^2}\,\sin\theta\,d\phi$, and $e^3=\sqrt{F}\,
(d\psi-2N\,\cos\theta\,d\phi)$. It has a non-trivial flux, and the
square of this form is
\be L_\2^{\pm\,2}=\fft{16}{(r\pm N)^4} \,. \ee
We will now break up further analysis for the cases of the
Taub-NUT and Taub-BOLT instantons.

\vspace{10pt} \underline{Taub-NUT}

In the BPS limit, $M=N$, for which the four-dimensional metric
(\ref{4Dnut}) is that of the Taub-NUT instanton. In this case,
$r\ge N$. The geometry goes from $\R^4$ to $C(S^2)\ltimes S^1$.

Only $L_\2^+$ is square integrable for $r\rightarrow N$. Neither
two-form is normalizable at large distance for any of the three
instantons. For $L_\2^+$, a regular solution is given by
\be H=1+\fft{m^2}{N(N+r)}\,. \ee
For $L_\2^-$, the unavoidably singular solution is given by
\be H=1+\fft{c}{r-N}-\fft{2m^2\,(r-3N)}{3(r-N)^3}\,. \ee
The geometry of this 5-brane interpolates from $M_{10}$ at short
distance to $M_6\times C(S^2)\ltimes S^1$ at large distance. This
solution preserves minimal supersymmetry \cite{cglptrans}.

\vspace{10pt} \underline{Taub-BOLT}

In this case, $M=\fft{5}{4}N$ and $r\ge 2N$. The geometry goes
from $\R^2\times S^2$ to $C(S^2)\ltimes S^1$.

Both $L_\2^{\pm}$ are square integrable for $r\rightarrow 2N$.
Regular solutions are given by
\be H=1+\fft{8m^2}{9N(N+r)}\,, \ee
and
\be H=1+\fft{8m^2}{N(r-N)}\,, \ee
for $L_\2^+$ and $L_\2^-$, respectively. Both of these 5-brane
geometries run from $M_8\times S^2$ to $M_6\times C(S^2)\ltimes
S^1$. $L_\2^+$ and $L_\2^-$ can be linearly superimposed to yield
a more general deformed 5-brane solution. However, as we
previously mentioned, the four-dimensional Taub-NUT instanton does
not admit a spin structure, though it may still admit a $Spin^C$
structure \cite{chamblinads}.

\vspace{10pt} \underline{Schwarzchild instanton}

For $N=0$, $r\ge 2M$. $L_\2^{\pm}$ are square integrable as
$r\rightarrow 2M$. This geometry goes from $\R^2\times S^2$ to
$\R^4$.

Both $L_\2^{\pm}$ have a regular solution given by
\be H=1+\fft{2m^2}{M\,r}\,. \ee
The corresponding 5-brane geometry go from $M_8\times S^2$ to
$M_{10}$.

\subsection{$S^1$-wrapped 5-brane}

The above resolution of the heterotic 5-brane requires the
presence of matter Yang-Mills fields which are absent in the type
II theories. However, a resolved type II 5-brane can be
constructed by wrapping worldvolume directions around the
transverse space. For example, a regular $S^2$-wrapped 5-brane was
obtained in \cite{s2} by lifting the four-dimensional $SU(2)$
gauged black hole \cite{su2bh}. This solution can apply for both
type II and heterotic 5-branes.

Another example of a regular 5-brane solution of both type II and
heterotic theories is the $S^1$-wrapped 5-brane given by
\cite{lvres}
\bea ds_{10}^2 &=& H^{-1/4}\,\Big( -dt^2+dx_1^2+\cdots +dx_4^2
+(dx_5+A_\1)^2\Big) +H^{3/4}ds_4^2\,,\nn\\
F_\3^{{\rm RR}} &=& \ast_4 dH-m\,L_\2\wedge (dx_5+A_\1)\,, \qquad
\phi=-\fft12\,\log H\,, \eea
where $dA_\1=m\,L_\2$. $L_\2$ is a harmonic 2-form on $ds_4^2$ and
$\ast_4$ is the Hodge dual with respect to $ds_4^2$. The equations
of motion are satisfied, provided that (\ref{eqnH5}) and $L_\2$ is
a self-dual two-form. Note that overlapping $S^1$-wrapped 5-branes
can also be resolved \cite{wrapped}.

We can also consider a rotating 5-brane, given by
\bea ds_{10}^2 &=& H^{-1/4}\,\Big( -(dt+A_\1)^2+
dx_1^2+\cdots +dx_5^2 \Big)+H^{3/4}ds_4^2\,,\nn\\
F_\3^{{\rm RR}} &=& \ast_4 dH-m\,L_\2\wedge (dt+A_\1)\,, \qquad
\phi=-\fft12\,\log H\,. \eea

For the Taub-NUT/BOLT metric given by (\ref{4Dnut}), the
computation of $L_\2$ and $H$ carry over from the deformed 5-brane
of the previous section. Since the Taub-BOLT and Schwarzchild
instantons both support two independent harmonic two-forms, these
can be superimposed to form a regular $T^2$-wrapped 5-brane. In
the case of the Schwarzchild instanton, for example, the resulting
$T^2$-wrapped 5-brane geometry goes from $M_4\times (\R^2\times
S^2)\ltimes T^2$ to $M_{10}$.

\section{Resolved D4/M5/NS5-branes}

The deformed D4-brane solution is given by
\bea ds_{10}^2 &=& H^{-3/8}\,dx_{\mu}^2+H^{5/8}\, ds_5^2\,,\nn\\
F_\4 &=& \ast\,dH\,,\qquad \phi=-\fft14 {\rm log} H\,,\nn\\
F_\2 &=& m\,L_\2\,,\qquad F_\3=m\,\ast\,L_\2\,, \eea
where $\ast$ is the Hodge dual with respect to $ds_5^2$. The
equations of motion are satisfied provided that $H$ is given by
(\ref{lapmod}), where $L_\2$ is a harmonic two-form on the
five-dimensional transverse space.

\subsection{On Schwarzchild instanton}

The $D=5$ Schwarzchild instanton metric is given by
\be ds_5^2=f\,dz^2+f^{-1}\,dr^2+r^2\,d\Omega_3^2\,, \ee
where
\be f=1-\fft{M}{r^2}\,, \ee
and $r\ge \sqrt{M}$. This geometry goes from $\R^2\times S^3$ to
$\R^5$.

This metric supports a harmonic two-form
\be L_\2=\fft{1}{r^3}\,dz\wedge dr\,, \ee
which is normalizable at large distance and square integrable at
short distance. The corresponding regular solution to $H$ is given
by
\be H=1+\fft{m^2}{4M\,r^2}\,. \ee
The geometry of this D4-brane smoothly interpolates from
$M_7\times S^3$ at short distance to $M_{10}$ at large distance.

The above deformed D4-brane can be lifted to eleven dimensions as
an $S^1$-wrapped M5-brane solution given by
\bea ds_{11}^2 &=& H^{-1/3}\,\Big( dx_{\mu}^2+
(dx_5+\fft{m}{2r^2}\,dz)^2\Big) +H^{2/3}\,ds_5^2\,,\nn\\
F_\4 &=& \ast dH+m\,\Big( dx_5+\fft{m}{2r^2}\,dz\Big) \wedge
\Omega_\3\,, \eea
where $\Omega_\3$ is the volume-form of the transverse $S^3$
corresponding to the metric $d\Omega_3^2$. The M5-brane geometry
runs from $M_5\times S^3\times \R^2\ltimes S^1$ to $M_{11}$.

We can now reduce back to ten dimensions along one of the
non-fibred spatial directions of the M5-brane. The result is the
$S^1$-wrapped D4-brane solution
\bea ds_{10}^2 &=& H^{-3/8}\,\Big( dx_{\mu}^2+
(dx_4+\fft{m}{2r^2}\,dz)^2\Big) +H^{2/3}\,ds_5^2\,,\nn\\
F_\4 &=& \ast dH+m\,(dx_4+\fft{m}{2r^2}\,dz)\wedge \Omega_\3\,.
\eea

Alternatively, we can express the transverse $S^3$ metric as a
fibre bundle over $S^2$:
\be d\Omega_3^2 = \fft14 (d\psi+\cos\theta\,d\phi)^2 +\fft14
d\Omega_2^2\,, \ee
where $d\Omega_2^2=d\theta^2+\sin^2\theta\,d\phi^2$. reducing over
the fibre bundle direction yields the type IIA $S^1$-wrapped
NS5-brane solution
\bea ds_{10}^2 &=& \Big( \fft{r}{2}\Big)^{1/4} \Big[ H^{-1/4}
\Big( dx_{\mu}^2+(dx_5+\fft{m}{2r^2}\,dz)^2\Big) +H^{3/4}
(f\,dz^2+f^{-1}\,dr^2+\fft{r^2}{4}\,d\Omega_2^2)\Big],\nn\\
F_\3 &=& \fft{r}{2}\ast dH+\fft{m}{8}\, \Big( dx_5+\fft{m}{2r^2}\,
dz\Big) \wedge \Omega_\2\,,\qquad F_\2 = \Omega_\2\,, \eea
where $\Omega_\2$ is the volume-form corresponding to the metric
$d\Omega_2^2$.

All of the above solutions are regular, since $r\ge \sqrt{M}$.

\subsection{On Schwarzchild on Taub-NUT}

A metric that resembles a $D=5$ Schwarzchild instanton
superimposed with a $D=4$ Taub-NUT instanton can be written as
\be ds_5^2=f\,dz^2+f^{-1}\,dr^2+2N(2N+M)\,W\,(d\psi+\cos\theta\,
d\phi)^2+(r^2-N^2)\,d\Omega_2^2\,, \ee
where
\be f=1-\fft{M}{r-N}\,,\qquad W=\fft{r-N}{r+N}\,, \ee
and $d\Omega_2^2=d\theta^2+\sin^2\theta\,d\phi^2$. Also, $r\ge
M+N$.

A harmonic two-form supported by this metric is given by
\be L_\2=\fft{1}{\sqrt{W}\,(r^2-N^2)}\,dz\wedge dr\,. \ee
which is normalizable at large distance and square integrable at
short distance. The corresponding regular solution to $H$ is given
by
\be H=1+\fft{m^2}{M\,N}\, {\rm log} \Big[
\fft{\sqrt{M(M+2N)(r^2-N^2)}+[(M+N)\,r-N^2]}{(\sqrt{M(M+2N)}+M+N)(r-N)}
\Big] \,. \ee

As in the previous section, we can obtain regular $S^1$-wrapped
M5, D4 and NS5-branes from the above solution.

\section{Conclusions}

We have constructed many regular $p$-brane solutions on
higher-dimensional generalizations of Taub-NUT and Taub-BOLT
instantons. These new solutions do not preserve any supersymmetry,
which serves to demonstrate that the resolution of brane
singularities works without the presence of supersymmetry. The
resulting geometries smoothly interpolate between two phases of
Minkowski spacetime of differing dimensionality. Since these new
solutions do not preserve any supersymmetry, the stability is not
assured. We leave this issue for future work.

\section*{ACKNOWLEDGMENT}

It is a pleasure to thank Hong L\"{u} for helpful correspondence.
Research is supported in part by DOE grant DE-FG01-00ER45832

\end{document}